\documentstyle[preprint,aps,epsf,amsmath,amssymb,graphicx]{revtex}

\begin{document}

\preprint{KVI-1552}
\draft
\tighten

\title{Covariant model for proton-proton bremsstrahlung: \\
       comparison with high-precision data}

\author{M.D. Cozma,$^a$, G.H. Martinus,$^a$ O. Scholten,$^a$ \\
        R.G.E. Timmermans,$^a$ J.A. Tjon$^{a,b}$}

\address{$^a\,$KVI, University of Groningen,
               Zernikelaan 25, 9747 AA Groningen, The Netherlands \\
         $^b\,$Institute for Theoretical Physics, University of Utrecht,
               Utrecht,The Netherlands}

\maketitle
\begin{abstract}
We compare a relativistic covariant model for proton-proton bremsstrahlung
with high-quality data from KVI. The agreement in large parts of phase
space is satisfactory. However, remarkably large discrepancies are
observed for specific kinematic regions. These failures are shown to
occur primarily when the final two-nucleon system has energies less than
about 15 MeV.
\end{abstract}

\subsection{Introduction}
The proton-proton bremsstrahlung ($pp\gamma$) process has
attracted significant attention over the years, both experimentally
and theoretically. In recent years, several new experiments have been
performed~\cite{Mic90,Bil98,Zlo98,Yas99,Hui99,Hui01}, inspiring many new
theoretical
investigations~\cite{Bro91,Nak92,Ede95,Mar97,Lio95,Kor95,Li98,Kon98}.
In particular,
a number of microscopic models have been developed to describe
the $pp\gamma$ process. Examples are the potential model of
Nakayama {\it et al}.~\cite{Nak92} and the covariant model of Martinus
{\it et al}.~\cite{Mar97}. The theoretical predictions of these models
could be compared to the $pp\gamma$ cross sections and analyzing
powers for the TRIUMF experiment at 280 MeV~\cite{Mic90}. The agreement
of theory with these TRIUMF data is rather good, especially for the
cross sections, provided that the experimental cross sections are
renormalized~\cite{Mic90}. Some outstanding discrepancies occur,
however, for certain asymmetric proton angles.

More recently, the first high-precision data from the KVI experiment
at 190 MeV became available~\cite{Hui99}, and many more data are
forthcoming~\cite{Hui01}. When comparing these data
with theory, a pronounced and undisputable discrepancy between theory and
experiment was observed in specific kinematic regions~\cite{Hui99}.
The size of the discrepancy between theory and experiment is disturbing,
since what primarily enters are the two-nucleon ($N\!N$) interaction and
the electromagnetic
coupling of the photon to the $N\!N$ system, both of which are believed
to be accurately known at this energy. The high precision of the new KVI
data allows one, in principle, to study smaller effects, like those arising
from negative-energy
states, the $\Delta$-isobar, and meson-exchange currents. It is therefore
important to identify the possible reasons for the discrepancies.

In this paper, we compare the covariant model of
Ref.~\cite{Mar97} with the KVI data available so far and analyze the
discrepancies. We demonstrate that the dominant contribution to $pp\gamma$
for the specific problematic kinematic regions results
when the $N\!N$ interaction is evaluated at energies below about 15 MeV,
and that, in fact, at least a major part of the problem resides in the
low-energy behavior of the $N\!N$ interaction models used.

This paper is organized as follows. First, we review briefly the
covariant model for $pp\gamma$ and compare it to the KVI data
in the problematic kinematic regions. Next, we demonstrate the sensitivity
of the bremsstrahlung cross sections in these regions to the properties
of the low-energy $N\!N$ interaction.
We then show that the problems can
be significantly alleviated accordingly by improving the description of
the interaction at low energies. We end by summarizing our conclusions.

\subsection{Covariant model for bremsstrahlung}
We first give a short review of the microscopic model of Martinus
{\it et al.} for $pp\gamma$~\cite{Mar97}. In this covariant
model the $N\!N$ $T$-matrix is obtained by solving the Bethe-Salpeter
equation for the two-nucleon system~\cite{Fle77} in the equal-time
approximation with the one-boson exchange (OBE) kernel of Ref.~\cite{Hum90}.
This OBE model for the $N\!N$ interaction was, at that time, fitted to
the Virginia Tech $np$ partial-wave solution~\cite{Arn87}, by adjusting
the meson-nucleon coupling constants and the form factor parameters, and
a reasonable agreement was obtained.

This covariant $N\!N$ $T$-matrix enters the model for $pp\gamma$, in which
a number of contributions can be distinguished. The most important ones
in the energy regime we consider here are the ``nucleonic'' contributions,
consisting of single-scattering terms, {\it i.e.} photon emission off the
external proton legs (see Fig.~1,~a-b), and the contribution commonly known
as rescattering (see Fig.~1,~c). The model is relativistic covariant and
therefore negative-energy states are included in a natural way. The relevance
of these negative-energy states is small for energies around 200 MeV. The
reason~\cite{Mar97} is that the contributions from the single-scattering
diagrams where the intermediate nucleons are in a negative-energy state
are canceled by similar contributions coming from the rescattering diagram.
This cancellation holds for terms up to order ${\mathcal O}(q)$, where $q$
is the photon momentum, and is a consequence of the soft-photon
theorem~\cite{Low58}.

Contributions from the $\Delta$-isobar and from magnetic meson-exchange
currents, containing in particular the $\omega\pi\gamma$ and $\rho\pi\gamma$
decay graphs, are also taken into account. These two-body current terms
are included in a perturbative way, since they are small in general.
The coupling constants of the photon to the various mesons were determined
from the radiative decay widths of the vector mesons.
As one can observe from Fig.~\ref{fig:2} below, at an energy of 190 MeV
the contribution of the two-body currents is small. These terms, however,
increase in size with energy and can become appreciable around the
pion-production threshold at 280 MeV and above.

This covariant bremsstrahlung model~\cite{Mar97} is theoretically well founded
and many of its ingredients have been tested in other calculations such 
as those for electron scattering on the deuteron~\cite{Hum90}. Therefore,
one did not expect {\em major} discrepancies with new experimental
$pp\gamma$ data at energies below 280 MeV.

\subsection{Comparison with the KVI data}
In the KVI experiment, $pp\gamma$ cross sections and analyzing powers
were measured for 190 MeV incoming proton energy, with the scattered
protons detected at small forward angles, and with the photon emitted
in the backward hemisphere~\cite{Hui99}. A typical example of the KVI
data~\cite{Hui99} in comparison with theory is shown is Fig.~\ref{fig:2}.
The data are plotted for two different kinematic situations, with $\theta_1$
and $\theta_2$ the fixed angles of the outgoing protons in the laboratory
frame, and as function of the polar angle of emission of the photon,
$\theta_\gamma$. The theoretical predictions are from the model
as published in Ref.~\cite{Mar97}.

 For the asymmetric proton angles $\theta_1=8^\circ$, $\theta_2=16^\circ$
(upper left panel)
the cross section shows a large discrepancy between theory and experiment
for values of $\theta_\gamma$ corresponding to the backward peak in the
cross section. For the symmetric proton angles $\theta_1=\theta_2=16^\circ$
(upper right panel), on the other hand, the cross section shows a much
better agreement between theory and data. The contribution of the
two-body currents is seen to be minor, and thus the discrepancy for
$\theta_1=8^\circ$, $\theta_2=16^\circ$ is unlikely to come from this
source. In the following, we focus therefore on the nucleonic contribution
as the cause of the problem.

Taking a closer look at the two kinematics presented in Fig. 2 reveals that
the one which poses problems ($\theta_1=8^\circ$, $\theta_2=16^\circ$) is
dominated by the contribution from the $^1S_0$ wave in the $N\!N$ $T$-matrix,
while for the second one for which the agreement is much better
($\theta_1=\theta_2=16^\circ$) the $^3P$ waves are as important
as the $^1S_0$ wave. The contribution of the rescattering diagram is
relatively small,
suggesting that the problem resides already at the level of the
single-scattering diagrams.

\begin{figure}[htbp]
\epsfxsize 7 cm
\centerline{\epsffile[240 410 450 570]{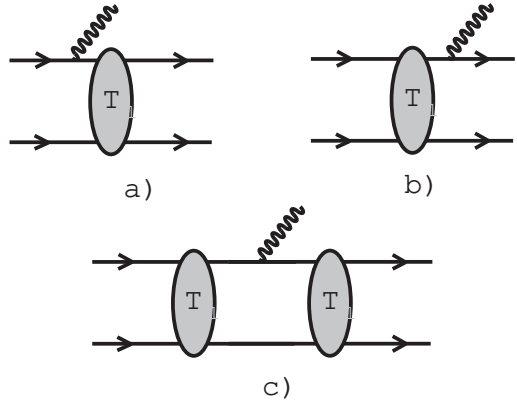}}
\caption[f1]{Single scattering, a) and b), and rescattering, c),
             contributions to $pp\gamma$. Analogous diagrams
             in which the lower proton radiates the photon are not shown.}
\label{fig:1}
\end{figure}

In the bremsstrahlung calculation the $N\!N$ $T$-matrix is evaluated at
different energies for each of the diagrams shown in Fig.~\ref{fig:1}. The
energy is lowest for the diagrams where the photon is emitted by one of the
incoming protons. The value of this energy is shown in Fig.~\ref{fig:3} as
a function of $\theta_\gamma$. For $\theta_1=8^\circ$, $\theta_2=16^\circ$,
the $T$-matrix is evaluated at 7 and 11 MeV kinetic energy for the
cases corresponding to the peaks in the bremsstrahlung cross sections,
{\it i.e.} at about $\theta_\gamma=20^\circ$ and $140^\circ$, respectively.
For the minimum in the cross section, around $\theta_\gamma=75^\circ$,
the $T$-matrix is evaluated at 24 MeV when the photon is emitted by one
of the incoming protons. In contrast, one observes that
for $\theta_1=\theta_2=16^\circ$, the $T$-matrix is evaluated
at more than 20 MeV for $\theta_\gamma=160^\circ$ and at more than 40
MeV for the minimum around $\theta_\gamma=75^\circ$; see the upper panel
of Fig.~3.
Thus, we conclude that theory and experiment agree well for
those situations where the $T$-matrix is evaluated above about 15 MeV, while
the discrepancies occur for the cases corresponding to a lower energy.

In order to demonstrate the sensitivity of the $pp\gamma$ cross section
to the low-energy $N\!N$ interaction, we plot in the lower panel of
Fig.~3 the cross section for $\theta_1=8^\circ$, $\theta_2=16^\circ$,
where we only include the $^1S_0$ wave, for three values of the scattering
length. Here we simply changed, for illustrative purpose only, the value
of $g^2_\varepsilon$, the coupling constant of the ``$\varepsilon$-meson''
in the OBE model.

In Table~1 we list for the two cases, $\theta_1=8^\circ$, $\theta_2=16^\circ$
and $\theta_1=\theta_2=16^\circ$, the cross
sections calculated by including in the $N\!N$ $T$-matrix only one of the
important partial waves, {\it viz.} the $^1S_0$ wave and the $^3P$ waves.
This is done for the value of $\theta_\gamma$ corresponding
to the backward peak or plateau in the cross section, {\em i.e.}
$\theta_\gamma=140^\circ$ in the first case and $\theta_\gamma=160^\circ$
in the second case, respectively. In both cases the first number listed
gives the cross sections with the $T$-matrix as used in Ref.~\cite{Mar97}
(the second number gives the corresponding value in the fit described below).
It is seen that the contribution of the $^1S_0$ wave is significantly larger
for the case $\theta_1=8^\circ$, $\theta_2=16^\circ$ compared to the case
$\theta_1=\theta_2=16^\circ$. By far the largest contribution
to this $^1S_0$ cross section comes from radiation off the incoming proton
legs. The $^3P$ wave cross sections arise mainly from radiation off the
outgoing proton legs. For $\theta_1=8^\circ$, $\theta_2=16^\circ$ emission
from the incoming proton legs dominates the bremsstrahlung cross section,
while for $\theta_1=\theta_2=16^\circ$ emission from initial
and final proton legs are comparable. The total cross sections listed in
the bottom row of Table~1 include also the rescattering contribution.

At the time of its construction, the OBE model of Ref.~\cite{Hum90} was
fitted to the $np$ phase-shift parameters of the Virginia Tech group, where
the interest of the fit was at the higher energies. In view of our findings
with the new high-precision $pp\gamma$ data, it becomes important
to improve the description at low energies. In order to investigate whether
at least the major part of the discrepancy between theory and the KVI data
can be removed, we have performed a preliminary refit with an emphasis of
obtaining better agreement with the Nijmegen $pp$ partial-wave analysis
of Ref.~\cite{Sto93}, at energies starting from 10 MeV up to 215 MeV.

In Fig.~4 we present the same calculation as in Fig.~2, but now for
the $N\!N$ model after the preliminary new fit. The description of
the $pp\gamma$ cross sections in the region of the backward peak for
$\theta_1=8^\circ$, $\theta_2=16^\circ$ has improved substantially. This
is mainly due to the improvement in the description of the $^1S_0$ wave
at low energies. Also the description of the $^3P$ partial waves has
improved. In Table~1 we now again list for the two kinematical situations
the contribution to the $pp\gamma$ cross section of the different
important partial waves. The new results should be compared to the
values obtained with the model as published in Ref.~\cite{Mar97},
and also listed in Table~1.

A preliminary investigation suggests that for those regions in phase
space where similar problems occurred as demonstrated here for the
case $\theta_1=8^\circ$, $\theta_2=16^\circ$, at least
a large part
of the problem is due to an inadequate description of the low-energy
behavior of the $N\!N$ $T$-matrix. 
Detailed results will be presented in Ref.~\cite{Coz01}.

Finally, it is worth pointing out that the sensitivity of the $pp\gamma$
cross sections to the low-energy $N\!N$ interaction raises the question
about the comparison between microscopic bremsstrahlung models of the type
discussed here and soft-photon descriptions. The soft-photon amplitude
of Ref.~\cite{Lio95} gives a remarkably good description of the experimental
$pp\gamma$ cross sections, including the data from KVI. In constructing
the soft-photon amplitude, a specific choice is made for the on-shell
points used in the calculation. For the kinematic situations discussed here,
this choice does not correspond to the $N\!N$ interaction at low energies,
and therefore, the relation to microscopic models is not obvious.

\subsection{Conclusions}
In conclusion, the $pp\gamma$ cross section at 190 MeV varies
strongly as a function of the angles of the protons. This can be traced
back to an increasing contribution of the $^1S_0$ wave for the regions
in phase space where the $N\!N$ interaction is probed at low energies.
For $\theta_1=8^\circ$, $\theta_2=16^\circ$, the $^1S_0$ wave dominates
strongly at the peak of the cross section. 
In contrast, for $\theta_1=\theta_2=16^\circ$, the
$^3P$ waves are as important as the $^1S_0$ wave.

The present study shows that the predicted cross sections in certain
kinematic regions are sensitive to the $N\!N$ interaction at low energy.
In all cases, the analyzing powers are affected less by changes in the
interaction.
We have in particular not included in these calculations the Coulomb
interaction, which plays an important role in the $pp$ system at low
energies. The results of such an analysis will be given in Ref.~\cite{Coz01}. 
Only a limited set of
KVI data has been published so far~\cite{Hui99}. With the complete
$pp\gamma$ data set~\cite{Hui01}, one may hope to be able to test in a
quantitative
way the validity of the various microscopic bremsstrahlung models.

\subsection*{Acknowledgments}
We acknowledge discussions with our experimental colleagues at KVI.
The research of R.G.E.T. was made possible by a fellowship of the
Royal Netherlands Academy of Arts and Sciences. We thank
M.C.M. Rentmeester for his help in setting up the codes for fitting
the $N\!N$ model.

\begin{table}
\begin{tabular}{c|cc|cc}
 & \multicolumn{2}{c|}{$\theta_1$, $\theta_2$, $\theta_\gamma$ $=$
                       $8^\circ$, $16^\circ$, $140^\circ$}
 & \multicolumn{2}{c} {$\theta_1$, $\theta_2$, $\theta_\gamma$ $=$
                       $16^\circ$, $16^\circ$, $160^\circ$} \\ \hline
  $^1S_0$  & 1.90 & $1.18$ & 0.39 & $0.20$ \\
  $^3P_0$  & 0.02 & $0.02$ & 0.01 & $0.02$ \\
  $^3P_1$  & 0.85 & $0.53$ & 0.82 & $0.51$ \\
  $^3P_2$  & 0.43 & $0.42$ & 0.53 & $0.47$ \\ \hline
  initial  & 2.31 & $1.53$ & 0.80 & $0.55$ \\
  final    & 0.93 & $0.68$ & 0.98 & $0.76$ \\ \hline
  total    & 2.40 & $1.98$ & 1.44 & $1.38$
\end{tabular}
\caption{Cross sections, in $\mu$b/sr$^2$rad, for different kinematics
         $\theta_1$, $\theta_2$, $\theta_\gamma$, split up in partial
         waves, radiation from initial and final proton legs, and the
         total. For each kinematics, the first column gives the value
         of the original fit~\protect\cite{Mar97}, while the second
         column gives the corresponding value after the refit.}
\label{tab:1}
\end{table}

\begin{figure}[htbp]
\epsfxsize 9 cm
\centerline{\epsffile[70 100 440 390]{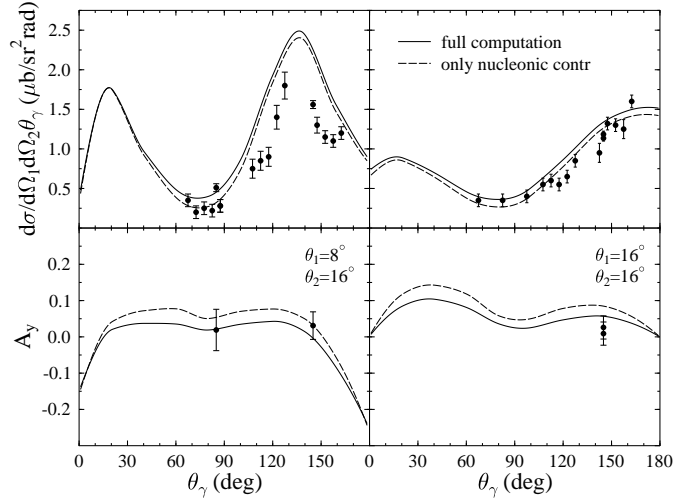}}
\caption[f2]{Bremsstrahlung cross sections (upper panels) and analyzing
             powers (lower panels) at 190 MeV incoming proton energy, and
             for proton angles $8^\circ$, $16^\circ$ (left panels)
             and $16^\circ$, $16^\circ$ (right panels). The solid curves
             show the results of the full model, including negative-energy
             states and two-body currents, while the nucleonic contribution
             is shown by the dashed lines. The KVI data (partly preliminary)
             are taken from Refs.~\cite{Hui99}.}
\label{fig:2}
\end{figure}

\begin{figure}[htbp]
\epsfxsize 8 cm
\centerline{\epsffile[90 240 480 690]{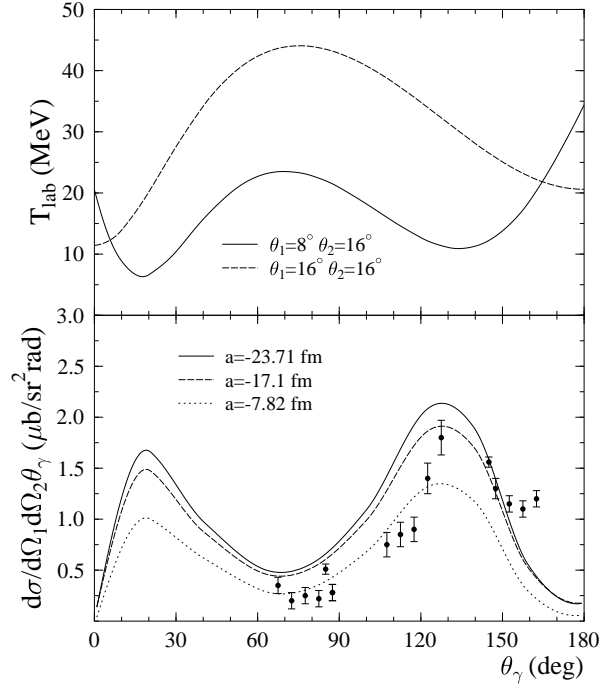}}
\caption[f3]{Upper panel: kinetic energy of the incoming proton at which
             the $N\!N$ $T$-matrix is evaluated for the two kinematics
             discussed in the text and in Fig.~2. Lower panel: bremsstrahlung
             cross sections at 190 MeV for three different scattering
             lengths for $\theta_1=8^\circ$, $\theta_2=16^\circ$. Only
             the contribution from the $^1S_0$ partial wave is included.}
\label{fig:3}
\end{figure}

\begin{figure}[htbp]
\epsfxsize 9 cm
\centerline{\epsffile[65 100 440 390]{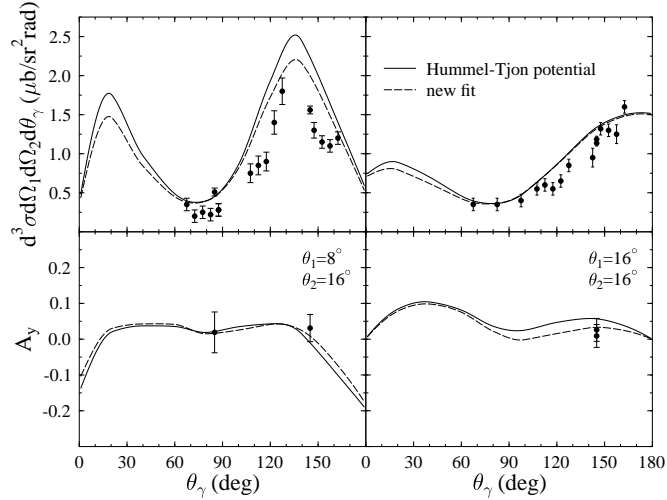}}
\caption[f4]{Bremsstrahlung cross sections and analyzing powers
             at 190 MeV, for the same kinematics as in Fig.~2.
             The solid (dashed) curves show the results before
             (after) the refit of the $N\!N$ model.}
\label{fig:4}
\end{figure}

\end{document}